\documentclass[prl, preprint, amsmath, superscriptaddress]{revtex4-1}
\usepackage{graphicx}
\usepackage{amsmath, amsthm, amssymb}
\usepackage{bm}

\newcommand{\beqn}{\begin{equation}}
\newcommand{\eeqn}{\end{equation}}

\begin{document}

\title{Selecting fast folding proteins by their rate of convergence.}

\author{Dmitry K. Gridnev}
\email[Electronic address: ]{gridnev@fias.uni-frankfurt.de}
\affiliation{FIAS, Routh-Moufang-Str. 1, 60438 Frankfurt, Germany}
\altaffiliation[On leave from]{St Petersburg State University, Uljanovskaja 1, 198504 St Petersburg, Russia}
\author{Pedro Ojeda-May}
\affiliation{Department of Chemistry and Chemical Biology,
Purdue University, Indianapolis, 402 N. Blackford st. LD 326 , Indiana, USA}
\author{Martin E. Garcia}
\email[Electronic address: ]{garcia@physik.uni-kassel.de}
\affiliation{Theoretische Physik, FB 10, and Center for Interdisciplinary Nanostructure Science (CINSaT),
Universit\"at Kassel,
Heinrich-Plett-Str. 40, 34132 Kassel, Germany}
\begin{abstract}

  We propose a general method for predicting potentially good folders
  from a given number of amino acid sequences.  Our approach is based
  on the calculation of the rate of convergence of each amino acid
  chain towards the native structure using only the very initial parts
  of the dynamical trajectories.  It does not require any preliminary
  knowledge of the native state and can be applied to different kinds
  of models, including atomistic descriptions.  We tested the method
  within both the lattice and off-lattice model frameworks and
  obtained several so far unknown good folders. The unbiased algorithm
  also allows to determine the optimal folding temperature and takes
  at least 3--4 orders of magnitude less time steps than those needed
  to compute folding times.
\end{abstract}


\maketitle



It is well-known, that most proteins fold rapidly and reliably to a unique
native state from any of a vast number of unfolded conformations
\cite{proteins,finkel}. One of the main problems in protein folding is
described by the so-called Levinthal paradox, which states that if the folding
pathway of a protein in the phase space would be governed by a random search
the time needed to locate the native state among all configurations would
exceed the age of the universe.  Nowadays, the consent answer to this paradox
is found in the designed energy landscape of a foldable protein, which
resembles a many-dimensional funnel, where moving along the free-energy
gradient narrows the accessible configuration space and guides to the unique
native structure, which lies at the bottom of the funnel
\cite{wolyn,jmol,wolyn2}. The funnel is also rough, giving rise to local
minima, which can act as traps during folding. In contrast to a designed
protein, a random amino acid chain will not fold to its global free-energy
minimum in times less than that needed to explore the configuration space
completely, the times, which are astronomically large \cite{wolyn}.

In this paper we call good folders those amino acid sequences, which exhibit a
protein-like behavior, {\em i.e.} those that fold into the unique native state
within a reasonable time. To find a way of characterizing good folders, like
typical motifs in the amino acid sequence or specific properties of the energy
landscape is of vital importance. A widely used criterion to characterize a
good folder is a pronounced energy gap between its global energy minimum and
the energies of configurations, which are structurally dissimilar to the
configuration of the global minimum (the native state) \cite{jmol,gap1,gap2}.
This energy gap ensures the ``thermodynamic stability'' and one finds a
correlation between the energy gap and the ability to fold into the global
minimum within a reasonable time. Yet, without knowing the native state, there
is still no good way to check whether a given amino acid sequence is a good
folder other than letting it dynamically evolve from various initial
conformations and checking if it does actually fold into a unique native state.
Due to an unknown folding time it may take very long before one could identify
some amino acid chain as a bad folder.

Many studies have been devoted to the search of determinants of a protein-like
system. Apart from the energy gap, one could mention the relation between the
folding and glass transition temperatures, see f. e.  \cite{cieplak}, the
collapse cooperativity \cite{klimov}, etc.  In this respect it is important to
understand how these features, which are characteristic to foldable proteins,
could help distinguish a good folder from a bad folder. Not always a clear
determinant of a good folder can serve as a criterion for selection of
protein-like aminoacid sequences.  It turns out that in order to do a fast
selection in most cases one needs to know the global minimum (native state)
from the beginning.  The energy gap clearly assumes the knowledge of the energy
in the native state. In \cite{microcanonical} the authors use the
microcanonical ensemble to distinguish good folders from bad folders but the
efficient procedure also requires the knowledge of the global energy minimum.
In simple Go--like models \cite{goreference}, where similar problems have been
posed (c.f. \cite{gin,go-mazzoni,epl}), the model space as a whole is biased by
the predetermined native state.  

In \cite{mazzoni} the authors propose an interesting idea to study the
fluctuations of the energy landscape curvature (this requires a smooth energy
surface).  This idea was tested on the off-lattice model with three amino
acids; the description of the model and some of the good folders can be found
in \cite{thirum2}. It  turns out that the averaged curvature of potential
energy $K_R := \nabla^2 U$ of a foldable protein suffers a dramatic enhancement
of the fluctuations in the vicinity of the folding temperature $T = T_f$. This
direction of research was further pursued in \cite{mazzoni2,go-mazzoni}.
Thereby, the preliminary knowledge of the native state is not necessary.
Successful selection of good folders in \cite{mazzoni,mazzoni2} was done from
only 6 sequences, which is too little to make a comparative analysis and to
judge on the effectiveness of the method. It is also important to note that the
curvature is averaged almost along the whole folding pathway, {i. e.} over the
whole folding time (the folding time can be found in \cite{thirum2}).
Sometimes the energy landscape is funneled towards several deep minima, and
since the approach in \cite{mazzoni,mazzoni2} is purely local, it is unclear
how one can distinguish good folders from bad folders in this case. Presumably,
this method works well when one compares a funneled and a totally frustrated
energy landscape, which was indeed the case in \cite{mazzoni,mazzoni2}.

In this paper using lattice and off--lattice models we investigate to which
extent the convergence of dynamical trajectories in configuration space on
early stages could serve a distinguishing criterion for a good folder.  We
emphasize that the knowledge of the native state is not required!  One can
illustrate the idea using a suitable analogy to convergence criteria for a
sequence of real numbers. On one hand, by definition, a sequence $A_n \in
\mathbb{R}$ for $n = 1, 2, \ldots$ converges if there exists $A_0 \in
\mathbb{R}$ such that for all $\varepsilon >0$ one can find $N$, which
guarantees that $|A_n - A_0|< \varepsilon$ holds for $n \geq N$.  Equivalently,
on the other hand, the sequence $A_n$ converges if for all $\varepsilon >0$ one
can find $N$ so that $|A_n - A_m|< \varepsilon$ holds for $n, m \geq N$.  In
the first case one needs to know the exact limit of a sequence (read native
state).  In the second case one does not have to know the limit of a sequence,
and similarly, it is not necessary to know the native state in our approach.

There are various ways to describe the dynamics of an amino acid chain in the
solvent (Langevin dynamics for atomistic models \cite{dyn1}, Monte Carlo (MC)
dynamics for lattice models \cite{hilhorst, gap2}, etc.). Generally, the time
development of the configuration can be written as $\mathfrak{C}(t) = g^t
\mathfrak{C}(0)$, where $\mathfrak{C}(0)$ is the initial configuration and
$g^t$ denotes the dynamical transformation, which depends on temperature and
has a probabilistic nature if it simulates how water molecules affect the amino
acid chain.

The effect of the folding funnel could also be expressed in terms of the
dynamical transformation, saying that if the dynamical transformation acts on
two arbitrary points in the configuration space then the ``distance'' between
them becomes contracted $d (\mathfrak{C}_1(t) , \mathfrak{C}_2 (t) ) <
d(\mathfrak{C}_1(0) , \mathfrak{C}_2 (0) )$, where $d$ stands for ``distance''
between configurations.  The time $t $ should surpass the minimal time required
for overcoming typical local traps in the folding funnel.  This expresses the
idea that if one considers a good folder in two randomly chosen initial
configurations and lets it dynamically propagate over a proper time, then there
should emerge structural similarities between two propagated yet initially
unrelated configurations.


Now imagine the following problem being posed: out of $K$ amino acid sequences
one has to sort out the best candidates for folding in some reasonable time.
The brute force solution to this problem would be to let each amino acid
sequence evolve according to the dynamics starting from various random initial
configurations and to check whether the dynamical trajectories reach the same
native conformation. This may be, however, extremely time consuming (especially
in the case of molecular dynamics simulations with water molecules included).
In addition, it is {\it a priori} unclear how long the dynamical simulation
must be run because the folding time is initially unknown.  Moreover, the
native contacts must not be necessarily known for an arbitrary sequence, which
prevents the application of go-type models.  In this paper we propose an
alternative solution to this problem based on comparing amino acid sequences
through their \textbf{rate of convergence}. To define the rate of convergence
for a given amino acid sequence $S$ we proceed as follows.

Suppose, the pairwise interaction between two monomers is $V_{ij} (\bm{r}_{ij}
)$, where $\bm{r}_{ij}$ is a relative coordinate between two monomers. Let us
extract the negative part of the potential function setting $W_{ij} (\bm{r} ):=
\max[0, -V_{ij}(\bm{r})]$ and define the magnitude of a contact between
aminoacids $i$ and $j$ as \begin{equation} \label{contacts} \overline V_{ij}
(\bm{r} ) := \frac{W_{ij} (\bm{r})} {\left(\max_{\bm{r}}[W_{ij} (\bm{r} )]
\right)} \quad \quad (\textrm{for $j\neq i-1, i, i+1$}), \end{equation} and
$\overline V_{ij} (\bm{r} ):= 0$ for $j= i-1, i, i+1$ (in the expression for
$\overline V_{ij} (\bm{r} )$ we exclude the bulk contributions from neighboring
monomers).  Clearly, $0 \leq \overline V_{ij} (\bm{r} ) \leq 1$.  Let
$\bm{r}^{(1)}_{ij}$ and $\bm{r}^{(2)}_{ij}$ denote $\bm{r}_{ij}$ in the
configurations $\mathfrak{C}_1$ and $\mathfrak{C}_2$ respectively. Then the
\textbf{overlap} between two configurations $\mathfrak{C}_1$ and
$\mathfrak{C}_2$ is defined as \begin{equation}\label{overlap}
\mathcal{O}(\mathfrak{C}_1 , \mathfrak{C}_2) = \sum_{i,j=1}^N \overline V_{ij}
(\bm{r}^{(1)}_{ij}) \overline V_{ij} (\bm{r}^{(2)}_{ij}), \end{equation} where
$N$ is the number of aminoacid molecules in the protein.  The overlap
introduces the topology in the space of configurations.  Note that the more
compact and structurally similar two configurations are the larger is the
overlap between them. Eqs.~(\ref{contacts}) and (\ref{overlap}) are quite
general and can be applied to any force field. As a particular case, for
lattice models $\overline V(\bm{r}_{ij}) = 1$ if the monomers are ``in
contact'' in the given configuration and zero otherwise.  For various
definitions of contact see, for example \cite{buchler,othercontacts}.

Next, let us fix some time scale $t_0$, which should be larger than the typical
time required for the dynamically evolving configurations to overcome local
minima on the energy surface. We then let a given amino acid chain dynamically
propagate over the time $t_0$ starting from two randomly chosen initial
configurations (self--avoiding random walks on the lattice) $\mathfrak{C}_{1}$
and $\mathfrak{C}_{2}$. The overlap between the resulting configurations
$g^{t_0} \mathfrak{C}_1 $ and $g^{t_0} \mathfrak{C}_2$ is then
$\mathcal{O}(g^{t_0} \mathfrak{C}_1 , g^{t_0} \mathfrak{C}_2)$. Sampling over
randomly chosen initial configurations $\mathfrak{C}_1$ and $\mathfrak{C}_2$ we
calculate the arithmetic mean of the overlaps, which we denote as $R (t_0,T) $
and call \textit{the rate of convergence} of the given amino acid sequence.
Here $T$ denotes the temperature (the dependence on $T$ is hidden in the
dynamical transformation). Below we would show that the rate of convergence $R
(t_0,T)$ can be used to select and design good folders. (In order to give a
proper dimension to the rate one could divide $R (t_0,T)$ by $t_0$; we do not
do this because this rescaling does not affect the results).  Let us  remark
that since the proteins coil into the native state from \textit{any} initial
configuration, we impose no restrictions on the domain of initial
configurations. 

Now we take the next step and construct the \textit{normalized rate of
convergence}.  For this purpose we first generate a large number of random
amino acid sequences and calculate $R(t_0 , T) $ for each sequence, where $t_0$
and $T$ are fixed time of evolution and temperature respectively. The
arithmetic mean of these values we denote as $R_{random} (t_0 , T) $.  This
quantity is the expectation value of the rate of convergence of a random
sequence depending on temperature and on the time scale $t_0$. The normalized
rate of convergence $R_N (t_0 , T)$ of an aminoacid sequence $S$ is then
defined as \begin{equation} \label{rateofconvergence} R_N(t_0 , T)  = R(t_0 ,
T)/ R_{random} (t_0 , T).  \end{equation} Let us remark that the values of
$R_{random} (t_0, T) $ can be tabulated so that  $R_N(t_0 , T) $ can be
determined with the same computational effort as $R(t_0 , T)$.

If an amino acid sequence has $R_N(t_0 , T) > 1$ then its rate of convergence
is larger than that of a random sequence; the converse is also true. The
normalized rate of convergence can be assigned to any amino acid sequence and
the larger $R_N (t_0,T)$ the better are the chances for this sequence to be a
good folder. Therefore, the best candidates for being a good folder from a
number of given amino acid sequences can be found through sorting the sequences
by their normalized rate of convergence.  The degree to which this sorting
algorithm is effective depends on how $t_0$, which is sufficient for proper
sorting, relates to the mean folding time. In the following we demonstrate that
the selection and design of good folders using the rate of convergence works
for both a standard lattice and an off-lattice models of proteins
\cite{jmol,gap1,tia}.

Although geometrically poor, the lattice model is protein-like in the sense
that lattice proteins fold to a unique native structure from an astronomically
large number of possible initial conformations and do so rapidly and
reproducibly. A random configuration is then a self avoiding random walk on the
cubic lattice.  The sequences are composed of 20 amino acids. Two monomers are
"in contact" if they occupy neighboring positions on the lattice but are not
sequence neighbors. The energy of two monomers in contact is calculated using
the $20 \times 20$ Miyazawa-Jernigan matrix (Table~VI in \cite{miya}).
The dynamic transformation $g^t$ is implemented through the Monte Carlo
dynamics \cite{tia} with move set including end moves, corner flips, and
crankshaft moves.

We have chosen a designed sequence \cite{seq} of 36 monomers S$_0$ =
SQKWLERGATRIADGDLPVNGTYFSCKIMENVHPLA.  The native state of S$_0$ has the energy
$E_{Nat} = -16.5$ in dimensionless $k_B T_{room}$ units, where $T_{room}$
stands for the room temperature \cite{miya}. At the folding temperature $T_f =
0.25$ (in Miyazawa-Jernigan dimensionless units) the configuration S$_0$ always
reaches its native state starting from any conformation and the mean folding
time (obtained by sampling $10^3$ self-avoiding random walks in initial
configurations) is $t_f = 1.5 \times 10^{6}$ steps.

In our calculations we have generated 800 sequences with a random amino acid
decomposition and the designed sequence S$_0$ was hidden among random sequences
as "a needle in a haystack".  For each amino acid sequence we calculated the
normalized rate of convergence and then sorted all sequences by the
corresponding value in descending order.  We computed $R_N(t_0,T_f )$, where
$T_f = 0.25$ is the folding temperature of S$_0$, over 500 randomly chosen
pairs of positions (conformations), starting with $t_0 = 50$ and repeated the
procedure incrementing each time $t_0$ by $50$. The initial conformations are
generated as self-avoiding random walks in the lattice.  We stress that for
each new time period the 800 random sequences were generated anew. We have
observed that further increase of the number of random sequences changes the
value  of $R_{random}(t_0 , T)$ by $\pm 1\%$ in the considered range of $t_0 ,
T$. Recall that these values can be obtained once with a high accuracy and then
tabulated for various values of $t_0 , T, N$, where $N$ is the number of
monomers. 

In general, for $t_0 \leq 150$ the designed sequence gets lost among other
random sequences, indicating that the time $t_0 \leq 150$ is insufficient for
overcoming local minima through potential barriers.  For $t_0 \geq 200$ the
sequence S$_0$ gets into the top ten, which makes us conclude that $t_0 \geq
200$ is sufficient for distinguishing the sequences by their ability to fold.
The dependence of normalized rate of convergence on the temperature $T$ for
fixed $t_0$ is also a relevant quantity. Remarkably, $R_N(t_0,T ) $ of S$_0$
peaks \textbf{exactly} at the folding temperature $T_f$, see Fig.~\ref{fig:Fig
1}.

In order to show that the rate of convergence can also be used to perform
sequence design we applied the algorithm to 5000 randomly generated amino acid
sequences having 36 monomers. The top 5 sequences turned out to be good
folders. We used $t_0 = 200$ and the sampling was done over $300$ pairs of
initial positions. The temperature was set to the folding temperature of the
designed sequence S$_0$, namely $T = T_f$.  Interestingly, the sequence S$_0$
occupied only the position 3.  The two top folders found correspond to the
sequences S$_1 =$ KWEEHEWGKDNLSDLHMHENEERFAQEQHNRDPQTD and S$_2 =$
NALCDDCSTEWCIPSMCCMCFEFIDFYKKKQQWRQM. The native states of S$_1$ and S$_2$ are
shown in Fig.~4. The energies of the native states are $E_{Nat} (S_1) = -16.88$
and $E_{Nat} (S_2) = -14.29$ respectively.  Note that $E_{Nat} (S_1)$ is even
lower than that of the previously known sequence S$_0$, despite the fact that
S$_1$ has the number of native contacts by 6 less than S$_0$ (note that the
structure of S$_0$ was specifically designed to maximize the number of native
contacts and 40 native contacts is the maximal reachable number for the
sequence length of 36 monomers).  Fig.~1 shows the normalized rate of
convergence for the sequences S$_0$ and S$_1$ as a function of temperature.  In
the given temperature range the normalized rates of convergence for S$_{1}$ is
larger than that of S$_0$. The same occurs for S$_{2}$ (not shown in Fig.~1).

Both newly found sequences S$_{1,2}$ have the folding temperature equal to
$T_f$ and their folding time is approximately 50 times longer than the folding
time of S$_0$. This is the fact which deserves a discussion: in spite of
S$_{1,2}$ having at all temperatures a better normalized rate of convergence
compared to S$_0$, their folding time is substantially longer. In \cite{seq}
one finds the procedure for the sequence design, where one fixes the target
conformation and finds the amino acid sequence, which minimizes the energy in
this conformation.  The target structure then becomes the native state for the
obtained good folder. The same design works also in the case of off--lattice
models \cite{clementi}.  The sequence design in our approach does not fix the
native conformation but rather fixes the target temperature. The obtained good
folders have the folding temperature equal to the target temperature! 

In addition, we applied our method to other sequences already designed by other
authors. For instance, for the sequences in Figs.~1,2 of Ref.  \cite{refsugg}
the method yields excellent results. In Fig.~1 we also plot the rate of
convergence versus temperature for the sequence S$_3 =
GYLGEIWKIMWAEMMKSWMSGWKGGEMGEWLKGIKG$ (Fig.~2 in \cite{refsugg}).  The curve
peaks exactly at the folding temperature.


As we have mentioned before, the rate of convergence $R(t_0,T)$ of a given
sequence is calculated by sampling over randomly chosen pairs of initial
conformations.  If one consider 100 pairs of random initial conformations then
the distribution of the overlaps for $t_0=300$ and $T=T_f$ is almost Gaussian
(as it should be in the perfect case according to the central limit theorem).


We now demonstrate that the method proposed here is also able to characterize
and design good folders in the more sophisticated off-lattice model of proteins
proposed by Clementi \textit{et al.} in \cite{clementi}.  In this force field
the interaction between amino acids $i$ and $j$ is given by \cite{erojas},

\begin{equation}
\label{eq:off-lattice}
V_{ij}= \delta_{i,j+1}a (r_{ij}-r_0)^2 + (1-\delta_{i,j+1})4\epsilon_{ij}
\left [ \left ( \frac{\sigma_{ij}}{r_{ij}} \right )^{12} - \left (
\frac{\sigma_{ij}}{r_{ij}} \right )^6 \right ],
\end{equation}
where $a=50$~\AA$^{-2}$ and $r_0=3.8$~\AA. The set of parameters $\epsilon$ and
$\sigma$ denote the minimum energy and the equilibrium distance for the
Lennard-Jones (LJ) part of the potential.  We considered $N_{conf}$ sequences
(with $N_{conf}$=100) of $N=30$ monomers.  To compute the time evolution $g^t$
of the monomers we used Monte Carlo dynamics.  The overlap between
configurations was computed using Eq.~(\ref{overlap}) and the rate of
convergence was obtained by averaging over $N_{conf}\times N_{conf}/2 =4950$
pairs of randomly chosen conformations, which were determined as follows.
First, we have chosen random positions for the monomers in the range [0:16] in
units of distance without any bias. Then, the so generated structures were
equilibrated during 2000 Monte Carlo steps, thus generating the starting
structural configurations.

We analyzed 6 sequences (see Table~\ref{tab:table1}) belonging to 3 different
polymer types according to the classification in \cite{erojas}. We considered 3
sequences of heteropolymer character (DHTP), labeled as SEQ1, SEQ2 and SEQ3, 2
sequences of random heteropolymers (RHTP) (SEQ4 and SEQ5) and the homopolymer
(SEQ6).  In general, heteropolymers designed following the procedure introduced
in \cite{erojas} have good chances to be protein-like, whereas for random
heteropolymers and for homopolymers one expects a rugged energy landscape and
consequently a bad folding behavior.

Note that SEQ1 has been shown to be a good folder, whereas SEQ4 and SEQ6 have
been previously characterized as bad folders \cite{erojas}.  The sequences
SEQ2, SEQ3 and SEQ5 generated by us in this work were not considered so far in
the literature.

The rate of convergence clearly allows one to
separate good folders from bad ones already at almost any step of the dynamical
simulation.  Fig. 2 shows the rate of convergence as a function of time for the 6 studied
sequences at fixed temperature. From the inset of Fig.~2 one can see that good folders can
be identified already after less than $10^4$ time steps, i.e., at an early
stage of the dynamical transformation $g^t$. At folding temperature our method allows
for a selection of good folders by computing trajectories {\it at least 3 to 4
orders of magnitude} smaller than those needed to compute the folding time. 

In Fig.~3 we show the temperature dependence of the normalized rate of
convergence $R_N(t_0, T)$ for the 6 sequences studied. The values of $R_{random} (t_0 , T) $ 
were computed using 100 random sequences; further increase of the number of random sequences 
changes the value  of $R_{random}(t_0 , T)$ by $\pm 2.5\%$ in the considered range of $t_0 ,
T$. Let us stress that one can get a better accuracy for $R_{random} (t_0 , T) $ using a larger number of random 
sequences; this does not affect the effectiveness of the method since for all models the  values $R_{random} (t_0 , T) $ 
can be tabulated after being calculated once.  

The normalized rate of
convergence was computed over 100 random sequences SEQ1, SEQ2,..., SEQ100, from
which SEQ1, SEQ2 and SEQ3 belonged to the DHTP model, SEQ6 was a HMP and the
rest of the sequences were random heteropolymers (RHTPs).  The different
functional dependence of good and bad folders is very clear. For good folders
$R_N(t_0, T)$ is larger than 1 at all temperatures and exhibits a well defined
maximum, whereas for bad folders $R_N(t_0, T) \simeq 1$ and practically does
not depend on temperature.

In order to investigate whether the temperature dependence of $ R_N(t_0, T)$ is
also physically relevant as in the case of the lattice model, we performed
Wang-Landau Monte Carlo simulations to calculate the specific heat curves of
the three good folders. Results are displayed in the low panel of Fig.~3.  The
specific heats of SEQ1, SEQ2 and SEQ3 show the typical peaked shape at the
folding temperatures $T_f$(SEQ$_i$), $i=1,2,3$, characteristic of protein-like
sequences.  By comparing the upper and lower panels of Fig.~3 one concludes
that from the position of the maxima of $R_N(t_0=10^7, T)$ one obtains a
reasonably good approximation to the folding temperatures. In order to obtain
smooth curves of $R_N $ vs T as those shown in Fig.~3  one has to take large
values of $t_0$.  From Fig.~3 it is clear that for each sequence $R_N(t_0, T)$
exhibits a broad maximum around $T_f$. Again, let us stress that the rate of
convergence is not only efficient in distinguishing good and bad folders but
also accurately predicts the suitable temperature range for a good folder.

Finally, we demonstrate that the new sequences SEQ2 and SEQ3, designed using
the method of the rate of convergence, are indeed foldable.  We computed the
average root mean square deviation
\begin{equation}
\label{Eq:rms}
\theta (t) =  \frac{1}{N_{conf}}\sum_{\nu=1}^{N_{conf}}\sqrt{\frac{2}{N(N-1)}
\sum_{i=1}^N \sum_{j>i}^N |\vec r^{Nat}_{ij,\nu} - \vec r_{ij,\nu}(t)|^2},
\end{equation}
where $r^{Nat}_i$ refers to the intermonomer distances in the native state and
$N_{conf}= 100$ to the number of initial conformations we average over. In
Fig.~4 we show the behavior of $\theta$, averaged over 100 independent trajectories,
as a function of $\log_{10}(t)$ for sequences SEQ1, SEQ2 and SEQ3. 
We can define the folding time as the time when $\theta$ approaches a certain
threshold value $\theta_{thr}$. We set $\theta_{thr} \sim 3.9~\AA$, which allows to estimate the
folding times as $t_{f}(SEQ1) = 4.5 \times 10^{6}$ time steps, $t_{f}(SEQ2)
= 6.6 \times 10^{5}$ time steps, and $t_{f}(SEQ3) = 2.4 \times 10^{7}$.

The three dimensional structures of some of the sequences designed in this work
using the rate of convergence 
are shown in Fig.~5. Note that the main conclusion of this paper, namely, that
the computational time required by the method of the rate of convergence is
many orders of magnitude less than the folding time remains valid even taking
into account that the definition of $R$ involves sampling over many different
initial conditions.  Such sampling operations can be run absolutely parallel on
as many different nodes as initial conditions one needs. Let us, however,
mention that the procedure presented here is, indeed, a good method to identify
potentially good folders, but it cannot serve as an ultimate measure of a good
folder.

The method of the rate of convergence developed in this paper is applicable in
all model frameworks which allow for dynamics, including accurate atomistic
descriptions.  Note that the rate of convergence $R$ can also be computed
basing on arbitrary definitions of overlap, different from Eqs. (1) and (2).
Moreover, it must not be restricted to the coordinate (structural) space. One
could, for instance, consider the overlap between strings containing property
factors \cite{kidera} or their Fourier components \cite{rackovsky}.

The authors express their gratitude to Dr. Guido Tiana for providing his
lattice-model dynamics software.


\begin{table}
\caption{\label{tab:table1} The six sequences studied in this
paper and their corresponding models. All the sequences have
$N=30$ monomers. The numbers in the second column denote the sequence
of amino acids in the peptide chain (using the same notation as in
Ref.~\onlinecite{clementi}.  }
\begin{ruledtabular}
\begin{tabular}{lcr}
Name& Sequence & Model\\
\hline
SEQ1 &311114442344312212224434333334  & DHTP\\
SEQ2 &341233331323231121112421234111  & DHTP\\
SEQ3 &443234423233421321132243424311  & DHTP\\
SEQ4 &414124323443321423324242141441  & RHTP\\
SEQ5 &444444444444444444444444444444  & RHTP\\
SEQ6 &321224314333113213344411112243  & HMP\\
\end{tabular}
\end{ruledtabular}
\end{table}



\begin{figure}
\includegraphics[width=.7\textwidth]{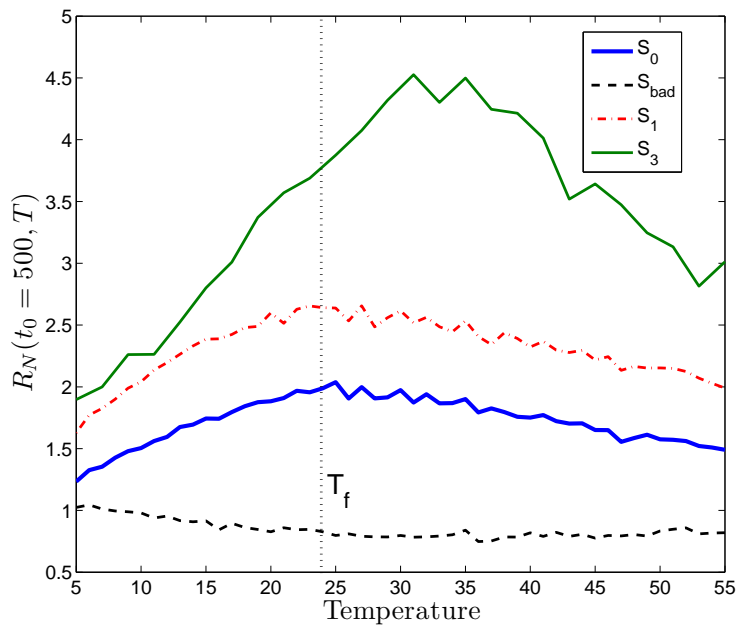}
\caption{(Color online). Thick solid line: the normalized rate of convergence versus temperature
   for the designed sequence S$_0$ for the time period $t_0 = 500$.
   Dash-dot and thin solid line : the same for the sequences S$_1$ and
   S$_3$ respectively. Note that the folding temperature of S$_3$ is
   approximately $1.2 T_f \simeq 30$ as can be seen from Figs.~9 (a,b)
   in \cite{refsugg}.  Dashed line: the normalized rate of convergence
   S$_{bad}$ of a typical bad folder (in this case a homopolymer). The
   vertical dotted line corresponds to the folding temperature of
   S$_0$.  The temperature is given in dimensionless Miyazawa-Jernigan
   units multiplied by 100 }\label{fig:Fig 1}
\end{figure}


\begin{figure}
\includegraphics[width=.7\textwidth]{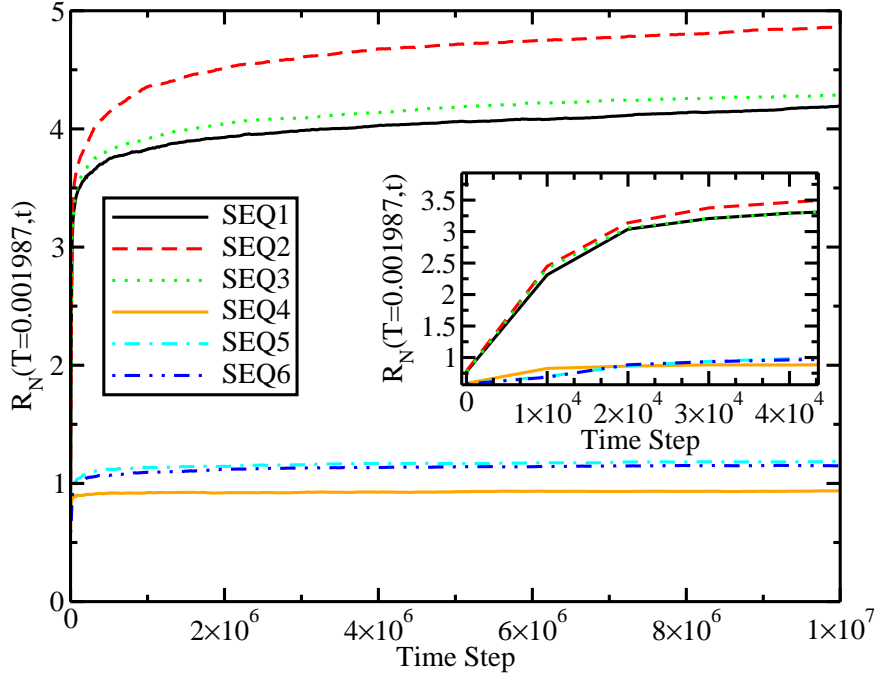}
\caption{(Color online). Normalized rate of convergence $R_N (t_0 , T)$ vs time step $t_0$ for fixed temperature $T= 0.001987 k_B^{-1}$
of the 6
 analyzed sequences in the off-lattice model (see the text). For each point, 
$R_N (t_0 , T)$ was calculated averaging over 100 conformation pairs. Inset:
 first stages of the time development of $R_N (t_0 , T)$.
The different behavior of good and bad folders is already evident.
}\label{fig:Fig2}
\end{figure}


\begin{figure}
\includegraphics[width=.7\textwidth]{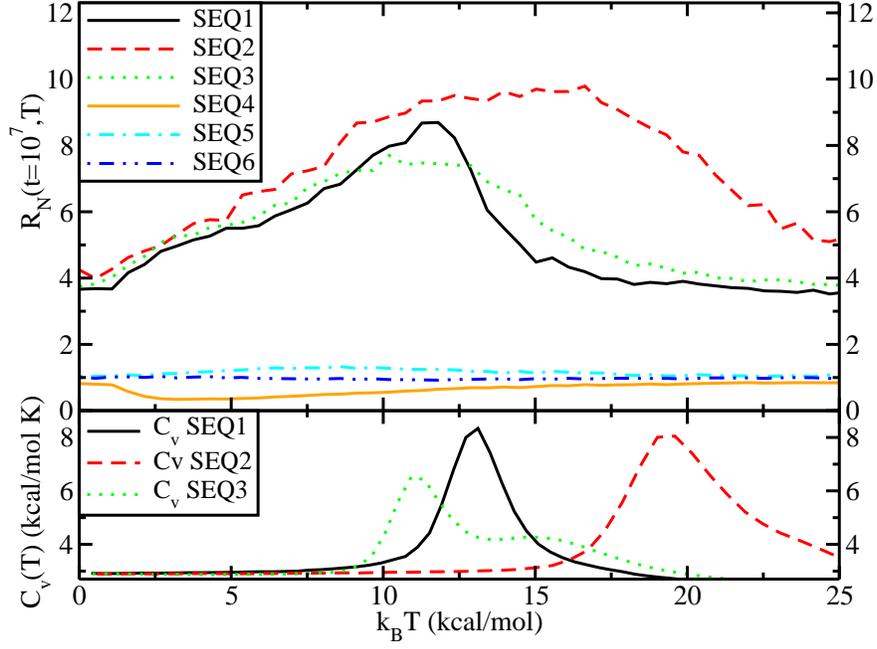}
\caption{(Color online). Upper panel: temperature dependence of the normalized rate of
 convergence  $R_N(t_0, T)$ for the 6 sequences considered
 within the off-lattice model. $t_0 = 10^7$ time steps. Lower panel:
specific heat curves of the sequences SEQ1, SEQ2 and SEQ3, characterized as
good folders by our method.
}\label{fig:Fig3}
\end{figure}

\begin{figure}
\includegraphics[width=.7\textwidth]{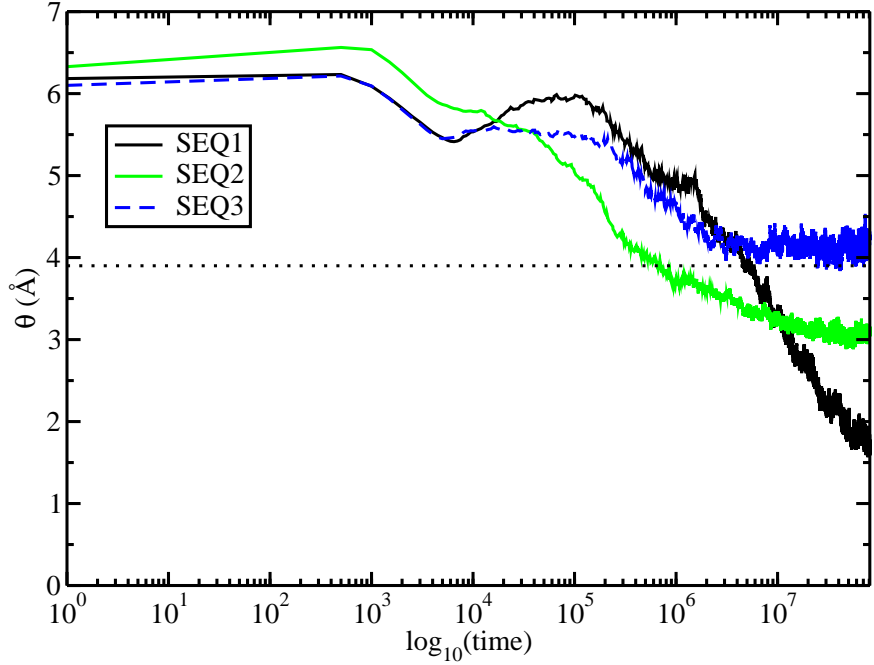}
\caption{(Color online). Time evolution of the root mean square deviation $\theta$ of sequences
SEQ1, SEQ2 and SEQ3 (with respect to their global minimum structures).
The value of $\theta$ was computed according to \ref{Eq:rms}.
Time axis is plotted in logarithmic scale. $\theta$ decays exponentially in time for the three sequences. 
Threshold value of 3.9~\AA~ is denoted by the dotted line. SEQ2 shows the
fastest folding  Monte Carlo dynamics followed by SEQ1 and SEQ3 (folding times are given in text).}
\label{fig:Fig4}
\end{figure}


\begin{figure}
\includegraphics[width=.8\textwidth]{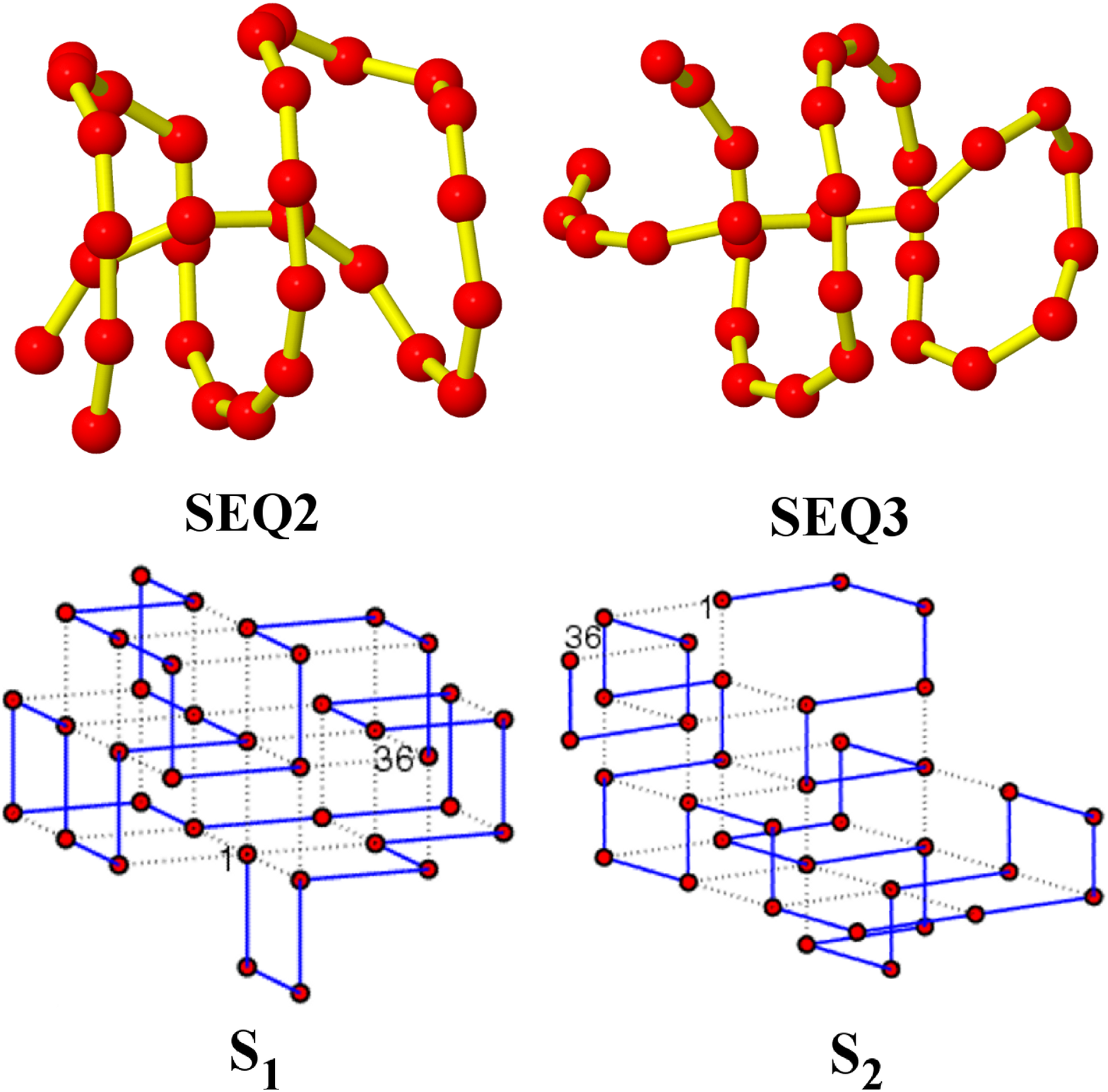}
\caption{(Color online). Native state conformations for some of the sequences
 designed using the rate--of--convergence method developed in this work. Lower
 panel: S$_2$ (left) and S$_3$ (right) obtained in the framework of the
 lattice model. Dotted lines connect those monomers that are in contact.
Upper panel: SEQ2 and SEQ3, designed within the off-lattice model.}
\label{fig:Fig5}
\end{figure}


\begin{thebibliography}{99}


\bibitem{proteins}
T. Creighton, Proteins Structure and Molecular Properties
(Freeman, New York, 1992).



\bibitem{finkel}
A. V. Finkelstein and O. B. Ptitsyn, {\em Protein Physics: A Course of Lectures}, (Academic Press, New York, 2002).


\bibitem{wolyn} R. Goldstein, Z. A. Luthey-Schulten, and P. Wolynes,
Proc. Natl. Acad. Sci. U.S.A. \textbf{89}, 4918  (1992).




\bibitem{jmol}
A. Sali, E. I. Shakhnovich, and M. Karplus, J. Mol. Biol.
\textbf{235}, 1614-1636 (1994).



\bibitem{wolyn2}J. Bryngelson, J. N. Onuchic, N. D. Socci, and P. Wolynes,
Proteins: Struct. Funct. Genetics \textbf{21}, 167 (1995).


\bibitem{gap1}
E. Shakhnovich and A. Gutin, Proc. Natl. Acad. Sci.
U.S.A. \textbf{90}, 7195 (1993);


\bibitem{gap2}
E. I. Shakhnovich, Phys. Rev. Lett. \textbf{72}, 3907 (1994).

\bibitem{cieplak} M. Cieplak, T. X. Hoang and M. S. Li, Phys.~Rev.~Lett. \textbf{83}, 1684  (1999)

\bibitem{klimov} D. K. Klimov and D. Thirumalai, Phys.~Rev.~Lett. \textbf{76}, 4070  (1996)

\bibitem{microcanonical} J. Hern\'{a}ndez-Rojas and J. M. Gomez Llorente, Phys.~Rev.~Lett. \textbf{100}, 258104 (2008)

\bibitem{goreference} V. Tozzini, Curr. Opin. Struct. Biol.  {\bf 15}, 144 (2005)

\bibitem{gin} B. C. Gin, J. P. Garrahan, P. L. Geissler, J. Mol. Biol. {\bf 392}, 1303 (2009).

\bibitem{go-mazzoni} J. Kim, T. Keyes, J. E. Straub, Phys.~Rev. E \textbf{79}, 030902͑R͒ (2009)

\bibitem{epl} L. Angelani and G. Ruocco, EPL \textbf{87} 18002 (2009) 


\bibitem{mazzoni}
L. N. Mazzoni and L. Casetti, Phys. Rev. Lett. \textbf{97}, 218104 (2006).

\bibitem{thirum2}
T. Veitshans, D. Klimov, and D. Thirumalai, Folding Des. \textbf{2}, 1 (͑1997͒).


\bibitem{mazzoni2}
L. N. Mazzoni and L. Casetti, Phys. Rev. E \textbf{77}, 051917 ͑(2008͒) 









\bibitem{dyn1}
M. K. Gilson, Proteins: Struct., Funct., Genet. \textbf{15}, 266 (1993).



\bibitem{hilhorst}
H. J. Hilhorst and J. M. Deutch, J.~Chem.~Phys.~{\bf 63}, 5153 (1975).


\bibitem{buchler}
M. Vendruscolo, R. Najmanovich, and E. Domany, Phys. Rev. Lett. \textbf{82}, 656 (1999).


\bibitem{othercontacts}
F. Birzele, J. E. Gewehr, G. Csaba, and R. Zimmer, Bioinformatics \textbf{23}, e205-e211 (2007);
I. Koch, \emph{ Ein graphentheoretischer Ansatz zum paarweisen und multiplen
Vergleich von Proteinstrukturen}, Wissenschaft und Technik Verlag, (1998).



\bibitem{tia}
R. A. Broglia, G. Tiana, H. E. Roman, E. Vigezzi and E. Shakhnovich, Phys. Rev. Lett. \textbf{82} 4727 (1999).



\bibitem{miya}
S. Miyazawa and R. Jernigan, Macromolecules \textbf{18}, 534 (1985).


\bibitem{seq}
V. Abkevich, A. Gutin, and E. I. Shakhnovich, Biochemistry
\textbf{33}, 10 026 (1994); G. Tiana, R. A. Broglia, H. E. Roman, E. Vigezzi, and E. I.
Shakhnovich, J. Chem. Phys. \textbf{108}, 757 (1998).



\bibitem{refsugg}
V. Abkevich, A. Gutin, and E. I. Shakhnovich, J. Mol. Biol. \textbf{252}, 460-471 (1995).



\bibitem{clementi}
C. Clementi, A. Maritan and J. Banavar, Phys. Rev. Lett. \textbf{81}, 3287 (1998).


\bibitem{erojas}
J. Hernandez-Rojas and J. M. Llorente, Phys. Rev. Lett. \textbf{100},258104 (2008).

\bibitem{kidera} A. Kidera, Y. Konishi, M. Oka, T. Ooi, and H. A. Scheraga. J Prot
Chem {\bf 4}, 23 (1985);  A. Kidera, Y. Konishi, T. Ooi, and H. A. Scheraga. J Prot Chem {\bf 4},
265 (1985).

\bibitem{rackovsky} S. Rackovsky, Phys. Rev. Lett {\bf 106}, 248101 (2011); Proc. Natl. Acad. Sci. U.S.A.
{\bf 107}, 8623 (2010).



\end{thebibliography}
\end{document}